\documentclass[aps,prl,groupedaddress,showpacs]{revtex4}
\usepackage{amsmath}
\usepackage{amsbsy}
\usepackage{graphicx}
\usepackage{upgreek}
\usepackage{amssymb}
\usepackage{epstopdf}
\usepackage{color}

\begin{document}
\title{Conserving optical currents: angular momentum and spin}%

\author{Michael Mazilu}
\email{mm17@st-andrews.ac.uk}
\affiliation{SUPA, School of Physics and Astronomy, University of St. Andrews, St. Andrews, KY16 9SS, UK}

\date{\today}

\begin{abstract}
Within the first quantisation of Maxwell's equations, we introduce the dynamic energy, linear momentum, angular momentum and optical spin of the electromagnetic fields. We show that these different quantities are conserved during the propagation of light in vacuum and are additive only for eigenmodes of the electromagnetic field.
\end{abstract}

\maketitle

\section{Introduction}

One conserving quantity that is transferred from the optical field to the scattering bodies is the angular momentum. In classical mechanics, the angular momentum is associated with the rotational state of a rigid body. An isolated physical system conserves its angular momentum meaning that its rotation continues ``in the same way'' for as long as no external torque acts upon it. The torque $\tau$ defines the amount of rotational force, with respect to a point, induced on a body by a force $F$. In a similar way, the electromagnetic angular momentum of a light beam can be defined as its ability to induce torque on an optically scattering object. There are two distinct mechanisms through which optical torque is transported by the electromagnetic wave. The first one corresponds to spin angular momentum that is conveyed by the polarisation of the light and interacts with birefringent materials for example. Indeed, a $\lambda/4$ optical waveplate, that converts linear polarised into circular polarised light, experiences a mechanical torque in the process. The second transfer mechanism has been termed orbital angular momentum and corresponds to the locally skewed motion of the energy flow of the light beam with respect to the average energy flow of the light beam. The orbital angular momentum of light can be observed in the case of optical micromanipulation of micro-particles in Laguerre-Gaussian beams, for example. Many different approaches have been developed to describe the optical spin and orbital momentum of light \cite{Enk:1994p9394,Barnett:2002p9290,Berry:2009p9030}.

Here, we use the operators introduced in quantum mechanics to deduce the linear, angular and spin momentum of electromagnetic fields. These operators act on the electromagnetic field solution of MaxwellÕs equation in the spinor Riemann-Silberstein form. For each operator, we can associate a conserving optical quantity and an electromagnetic current. The simultaneous conservation of all these quantities cannot be fulfilled, in an absorption or scattering process, except for optical eigen-modes of the associated operators \cite{Mazilu:2009p9267}. This, in turn, leads the formal first quantisation of MaxwellÕs equations. The developed formalism is applied to deduce the orbital and spin eigen-modes of light.

\section{Theory}

The starting point of the approach are Maxwell's equations in vacuum:
\begin{eqnarray}\label{max1}
\nabla\cdot{\bf E}=0 & \;\;\;\;\;\;& \nabla\times{\bf E} = - \frac{1}{c} \partial_t {\bf H} \nonumber \\
\nabla\cdot{\bf H}=0 & \;\;\;\;\;\;& \nabla\times{\bf H} =  \frac{1}{c}\partial_t {\bf E} 
\end{eqnarray}
where $\bf E$ and $\bf H$ are the electric and magnetic vector fields and where $c$ is the speed of light. 
Here, we consider a complex valued analytic representation of the fields i.e. the real and imaginary parts of the fields are linked via a time-domain Hilbert transform. 
Consequently, the spectrum of the complex valued electric and magnetic has no negative frequency components. 
Further, we chose the units of the electric and magnetic fields such that the energy density of a monochromatic wave is 
$\hbar \omega ({\bf E}^*\cdot {\bf E}+{\bf H}^*\cdot {\bf H})$ where the time dependance of the fields is given 
by the factor $\exp(i\omega t)$ and where $\omega$ is the optical frequency. 
In this units, the quadratic quantity $({\bf E}^*\cdot {\bf E}+{\bf H}^*\cdot {\bf H})$ can be seen as the local density of photons.  
We note that, the choice of the analytic representation simplifies the definition of the optical vector spin, while the choice of units clarifies its relationship with the quantum mechnical analogue. 

The complex electric and magnetic fields can each be decomposed into real and imaginary parts as ${\bf E}={\bf E}_r+i{\bf E}_i$ and ${\bf H}={\bf H}_r+i{\bf H}_i$.
Using this decomposition, we can introduce two Riemann-Silverstein (RS) equations equivalent with the complex Maxwell's equations (\ref{max1}):
\begin{eqnarray}
\nabla\cdot{\bf F}_r=0 & \;\;\;\;\;\;& \nabla\times{\bf F}_r = \frac{i}{c} \partial_t {\bf F}_r \nonumber \\
\nabla\cdot{\bf F}_i=0 & \;\;\;\;\;\;& \nabla\times{\bf F}_i = \frac{i}{c} \partial_t {\bf F}_i 
\end{eqnarray}
where the complex RS vector fields are defined by $ \mathbf{F}_r={\bf E}_r+i{\bf H}_r$ and  $ \mathbf{F}_i={\bf E}_i+i{\bf H}_i$. The introduction of the second ancillary RS equation is one of the main results of this paper. Indeed, this additional equation makes it possible to represent the polarization state the electromagnetic field without any additional operations on the fields \cite{Kaiser:2004p413}.

For convenience, these two RS equations can be combined into a single equation using the spinor notation $\boldsymbol F=({\bf F}_r, {\bf F}_i)$ which reads
\begin{eqnarray}\label{maxs}
\nabla\cdot \boldsymbol F =0 & \;\;\;\;\;\;& \nabla\times \boldsymbol F = \frac{i}{c} \partial_t \boldsymbol F. 
\end{eqnarray}
In the following, we treat the spinor field $\boldsymbol F$ as a two component vector.

Using the spinor or directly the RS field notation, we can describe the interference conservation relation 
\begin{eqnarray}\label{inter:cons}
0=\nabla\cdot {\bf j}_{12} +\partial_t \rho_{12}
\end{eqnarray}
where ${\bf j}_{12}$ and $\rho_{12}$ are, respectively, the complex interference current and its associated density defined by:
 \begin{eqnarray}\label{inter:def}
{\bf j}_{12}=&-i c\boldsymbol F_2^*\times \boldsymbol F_1&=-i c{\bf F}^*_{2r}\times {\bf F}_{1r}-i c{\bf F}^*_{2i}\times {\bf F}_{1i}\nonumber \\
\rho_{12}= &\boldsymbol F_2^* \cdot \boldsymbol F_1 &= {\bf F}^*_{2r}\cdot {\bf F}_{1r}+{\bf F}^*_{2i}\cdot {\bf F}_{1i}
 \end{eqnarray}
for any two general electromagnetic solutions, $\boldsymbol F_1$ and $\boldsymbol F_2$, of Maxwell's equations. These equations are also associated with a further set of three equations describing the conservation of the interference current together with its flux. 

\newlength{\gnat}
\settowidth{\gnat}{SM}
\begin{table}[htb]
\begin{tabular}{|c|c|c|c|}
\hline
Operator & $\cal T$  & $\rho_{12}(\cal T)=\boldsymbol F^* \cdot \cal T \boldsymbol F $ & ${\bf j}_{12}({\cal T})=-i c\boldsymbol F^*\times \cal T \boldsymbol F $ \\ \hline
$\rho$& $\boldsymbol\sigma_0$ & $  \mathbf{E}^*\cdot\mathbf{E}+ \mathbf{H}^*\cdot\mathbf{H}$ & $ c (\mathbf{E}^*\times\mathbf{H}- \mathbf{H}^*\times\mathbf{E})$\\
$\cal E$ & $i\hbar\partial_t \boldsymbol\sigma_0$ & $\mathbf{E}^*\cdot(i\hbar\partial_t)\mathbf{E}+ \mathbf{H}^*\cdot(i\hbar\partial_t)\mathbf{H}  $ & $ c (\mathbf{E}^*\times(i\hbar\partial_t)\mathbf{H}- \mathbf{H}^*\times(i\hbar\partial_t)\mathbf{E})$\\
$P_x$&  $-i \hbar \partial_x \boldsymbol\sigma_0$  &  $-\mathbf{E}^*\cdot(i\hbar\partial_x)\mathbf{E}- \mathbf{H}^*\cdot(i\hbar\partial_x)\mathbf{H}  $ & $ -c (\mathbf{E}^*\times(i\hbar\partial_x)\mathbf{H}- \mathbf{H}^*\times(i\hbar\partial_x)\mathbf{E})$\\
$P_y$&  $-i \hbar \partial_y \boldsymbol\sigma_0$  &  $-\mathbf{E}^*\cdot(i\hbar\partial_y)\mathbf{E}- \mathbf{H}^*\cdot(i\hbar\partial_y)\mathbf{H}  $ & $ -c (\mathbf{E}^*\times(i\hbar\partial_y)\mathbf{H}- \mathbf{H}^*\times(i\hbar\partial_y)\mathbf{E})$\\
$P_z$&  $-i \hbar \partial_z \boldsymbol\sigma_0$  &  $-\mathbf{E}^*\cdot(i\hbar\partial_z)\mathbf{E}- \mathbf{H}^*\cdot(i\hbar\partial_z)\mathbf{H}  $ & $ -c (\mathbf{E}^*\times(i\hbar\partial_z)\mathbf{H}- \mathbf{H}^*\times(i\hbar\partial_z)\mathbf{E})$\\
$L_x$ &$-(i \hbar\mathbf r\times \nabla)_x\boldsymbol\sigma_0$ & $  -\mathbf{E}^*\cdot(i \hbar\mathbf r\times \nabla)_x\mathbf{E}- \mathbf{H}^*\cdot (i \hbar\mathbf r\times \nabla)_x\mathbf{H} $ & $- c (\mathbf{E}^*\times(i \hbar\mathbf r\times \nabla)_x\mathbf{H}- \mathbf{H}^*\times(i \hbar\mathbf r\times \nabla)_x\mathbf{E})$\\
$L_y$ &$-(i \hbar\mathbf r\times \nabla)_y\boldsymbol\sigma_0$ & $  -\mathbf{E}^*\cdot(i \hbar\mathbf r\times \nabla)_y\mathbf{E}- \mathbf{H}^*\cdot (i \hbar\mathbf r\times \nabla)_y\mathbf{H} $ & $ -c (\mathbf{E}^*\times(i \hbar\mathbf r\times \nabla)_y\mathbf{H}- \mathbf{H}^*\times(i \hbar\mathbf r\times \nabla)_y\mathbf{E})$\\
$L_z$ &$-(i \hbar\mathbf r\times \nabla)_z\boldsymbol\sigma_0$ & $  -\mathbf{E}^*\cdot(i \hbar\mathbf r\times \nabla)_z\mathbf{E}- \mathbf{H}^*\cdot (i \hbar\mathbf r\times \nabla)_z\mathbf{H} $ & $ -c (\mathbf{E}^*\times(i \hbar\mathbf r\times \nabla)_z\mathbf{H}- \mathbf{H}^*\times(i \hbar\mathbf r\times \nabla)_z\mathbf{E})$\\
$S_0$ & $\hbar\boldsymbol\sigma_0$ &$ \hbar(\mathbf{E}^*\cdot\mathbf{E}+ \mathbf{H}^*\cdot\mathbf{H})$ & $ \hbar c (\mathbf{E}^*\times\mathbf{H}- \mathbf{H}^*\times\mathbf{E})$\\
$S_1$ & $\hbar\boldsymbol\sigma_1$ &$ 2\hbar( {\bf E}_i\cdot{\bf  E}_r+  {\bf H}_i\cdot{\bf  H}_r)$ &  $ 2\hbar c (\mathbf{E}_r\times\mathbf{H}_i+ \mathbf{E}_i\times\mathbf{H}_r)$\\
$S_2$&$\hbar\boldsymbol\sigma_2$ &$ 2\hbar({\bf E}_i\cdot{\bf  H}_r -{\bf E}_r\cdot{\bf  H}_i  )$ &  $ 2\hbar c (\mathbf{E}_r\times\mathbf{E}_i+ \mathbf{H}_r\times\mathbf{H}_i)$\\
$S_3$& $\hbar\boldsymbol\sigma_3$&$ \hbar({\bf E}_r\cdot{\bf  E}_r-{\bf E}_i\cdot{\bf  E}_i - {\bf H}_i\cdot{\bf  H}_i+ {\bf H}_r\cdot{\bf  H}_r)$ & $2 \hbar c (\mathbf{E}_r\times\mathbf{H}_r+ \mathbf{H}_i\times\mathbf{E}_i)$\\
\hline
\end{tabular}
\caption{Conserving quadratic quantities corresponding to the following operators leaving Maxwell's equations invariant: Photon Density ($\rho$), Energy Density ($\cal E$), Linear Momentum (${\bf P}=(P_x,P_y,P_z)$), Orbital momentum (${\bf L}=(L_x,L_y,L_z)$), Spin momentum $(S_0,S_1,S_2,S_3)$. For vector quantities we use the index $k$ taking alternatively the three coordinates $(x,y,z)$. }
\label{tab}
\end{table}

Noether's theorem states that any field transformation, $\cal T$, leaving  Maxwell's equation  (\ref{maxs}) invariant is associated with a conserving current. Considering a solution $\boldsymbol F$ of equation  (\ref{maxs}) one can define the two fields $\boldsymbol F_1= \cal T \boldsymbol F$ and $\boldsymbol F_2= \boldsymbol F$. These fields together with relation (\ref{inter:cons}) define the conserving current and density associated with the field transformation $\cal T$. Table \ref{tab} summarises the case of multiple field transformations $\cal T$ that lead to the definition of the energy, linear and angular momentum. The three spin momentum components are defined using the three Pauli matrices: 
$$ 
\boldsymbol\sigma_1 =\left(\begin{array}{cc}0 & 1 \\1 & 0\end{array}\right) \;\;\;\;\;\; 
\boldsymbol\sigma_2 =\left(\begin{array}{cc}0 & i \\-i & 0\end{array}\right) \;\;\;\;\;\;  
\boldsymbol\sigma_3 =\left(\begin{array}{cc}1 & 0 \\0 & -1\end{array}\right)
$$
and the identity matrix $\boldsymbol\sigma_0$. The spin operators introduced through this procedure are similar to the Stokes operators defined via the direct Quantum Mechanical approach \cite{Schnabel:2003p9392,Shumovsky:1998p9393}.

We remark that the rotation of the spinor base around the unit vector ${\bf u}=(u_1,u_2,u_3)$ with an angle $\alpha$ is defined by the operator 
\begin{equation}
{\bf R}=\exp(i(u_1 \boldsymbol\sigma_1+u_2 \boldsymbol\sigma_2+u_3 \boldsymbol\sigma_3) \alpha/2).
\end{equation}
This rotation implies  the rotation of the 3-D vector $(S_1,S_2,S_3)$ around $\bf u$ with the angle $\alpha$. Here, we remark that for the spatial rotation to correspond to a spinor base rotation we need to define an absolute base system.  

Further, we observe that the four components of the spin have the following relationship
\begin{equation}
S_1^2+S_2^2+S_3^2=S_0^2
\end{equation}
equivalent to relationship between the Stokes parameters for monochromatic waves. The difference between this two approaches lies in the base used to represent the Stokes parameters and the the one used for the optical spin. 
Further, the four spin components form a 4-vector provided the spinor field is transformed appropriately. For example, moving along the $z$ axis with a velocity $v_z$ will imply the spinor operator ${\bf R}_L=\exp( \boldsymbol\sigma_z \theta/2)$ where the rapidity $\theta$ is defined by  $\tanh(\theta)=v_z/c$. In this case the 4-vector $(S_0,S_1,S_2,S_3)$ transforms using the corresponding Lorentz transformation. 

For each operator in table \ref{tab} one can define the total conserving quantity as
\begin{equation}
<\boldsymbol F^* \cdot {\cal T} \boldsymbol F>=\int dx^3 \boldsymbol F^* \cdot {\cal T} \boldsymbol F
\end{equation}
which, because of the the conservation relation (\ref{inter:cons}), remain constant. For this quantity to the additive in a linear superposition of two fields $\boldsymbol F_1$ and $\boldsymbol F_2$ following relationship needs to be fulfilled
\begin{equation}
<(\boldsymbol F_1+\boldsymbol F_2)^* \cdot {\cal T} (\boldsymbol F_1+\boldsymbol F_2)>=<\boldsymbol F_1^* \cdot {\cal T }\boldsymbol F_1>+<\boldsymbol F_2^* \cdot {\cal T}. \boldsymbol F_2>
\end{equation}
This property ensures that in the creation, annihilation or amplitude change of one field, the corresponding conserving quantity change accordingly. For this property to be fulfilled, the field solutions $\boldsymbol F_1$ and $\boldsymbol F_2$ need be eigenfunctions of the operator $\cal T$
\begin{eqnarray}
\lambda_1\boldsymbol F_1&=& {\cal T} \boldsymbol F_1\nonumber \\
\lambda_2\boldsymbol F_2&=&{\cal T} \boldsymbol F_2.
\end{eqnarray}
In effect, for Hermitian symmetry operators $\cal T$, eigenfunction corresponding to  different eigenvalues  are orthogonal and as such will not interfere in a linear superposition. This explains how the quadratic measures corresponding to the conserving quantity are additive in a linear superposition of the fields. This property can only be extended to a number of Hermitian operators for which one can define common eigenfunctions.  Hermitian operators  that do not commute, such as the different spin operators, lead to conserving quantities that are not additive in any field superposition. The spin and orbit operators commute with each other. This makes the quantities associated with this operators appropriate to describe the solutions of the electromagnetic fields. In this case, the orbital angular momentum eigenfunctions
\begin{equation}
q\hbar    \boldsymbol F= -(i \hbar\mathbf r\times \nabla)_z \boldsymbol F
\end{equation}
 are Bessel beams propagating in the $z$ direction with  $q$  the topological charge. One can also verify that the spin operator eigenfunctions are given by linear and circular polarised light. 

\begin{figure}[htbp] 
   \centering
   \includegraphics[width=8cm]{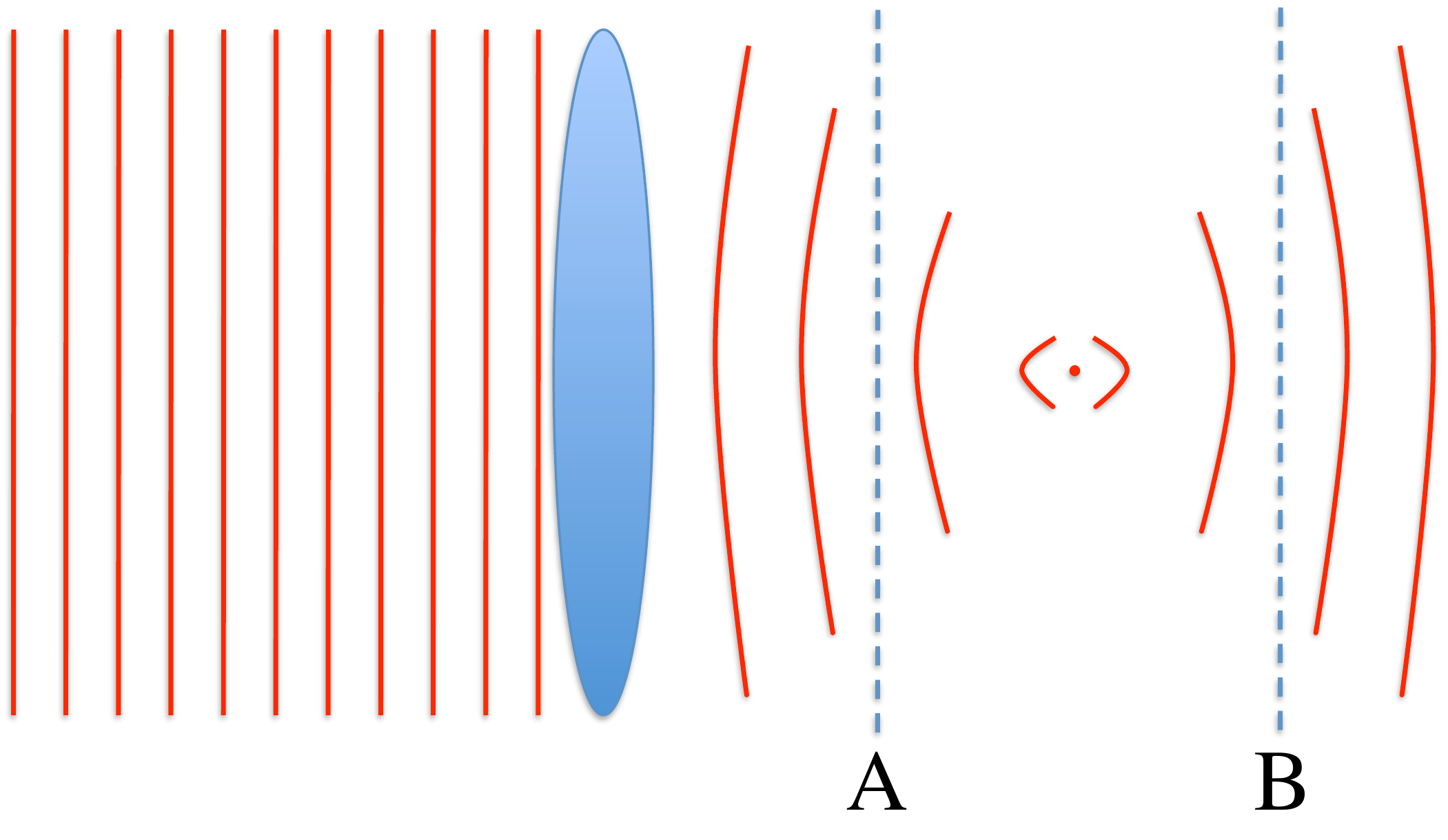} 
\caption{Tightly focussed light field. The dotted plane A and B correspond to the integration planes considered here. }
\label{fig:1}
\end{figure}

\section{Application}

An important question arising when studying the spin and orbital angular momentum of light is the condition in which transfer occurs between these quantities. The formalism and operators presented here are applicable as well in the paraxial regime as in the case of tightly focussed beams where the paraxial approach fails.  This latter case has been put forward as a mechanism to link orbital angular momentum to spin angular momentum\cite{Nieminen:2008p9066,Zhao:2007p8338}. Figure \ref{fig:1} shows the principle of the configuration considered. The light field is tightly focussed and we evaluate the flux of spin and orbital angular momentum through planes A and B. Here we consider the case of monochromatic light fields. In this case, the conservation relationship \ref{inter:cons} simplifies to  
\begin{eqnarray}\label{div}
0=\nabla\cdot {\bf j}_{12}. 
\end{eqnarray}
Taking into account that the fields at infinity are zero, this leads to the following integral relations between the flux in plane A and B
\begin{eqnarray}
\int_A {\bf j}_{12}(\hbar\boldsymbol\sigma_k )\cdot {\bf u}_z ds =\int_B {\bf j}_{12}(\hbar\boldsymbol\sigma_k )\cdot {\bf u}_z ds \\
\int_A {\bf j}_{12}(-(i \hbar\mathbf r\times \nabla)_k )\cdot {\bf u}_z ds =\int_B {\bf j}_{12}(-(i \hbar\mathbf r\times \nabla)_k )\cdot {\bf u}_z ds. 
\end{eqnarray}
These relations show that in vacuum (or other homogenous media), no transfer between the orbital and the spin angular momentum is possible. This approach is not only applicable in the case of tightly focussed beams but more generally in for any beams. This contrasts with the experimental observations by Zhao et al. \cite{Zhao:2007p8338} which might be explained when rigourously considering the effect of the different elements of the optical setup.  Further, strict conservation remains true when the surface of integration is changed to a sphere or any other surface provided the flux is projected on the normal of the considered surface. It is this projection that, when no taken into account, leads to the erroneous conclusion that momentum might transfer \cite{Nieminen:2008p9066}.   

\section{Conclusion}
The first quantisation of light brings Maxwell's equations to par with Schr\"odinger's equations. Formally, both equations are governed by the same symmetry operators leading conserving quantities having the same physical meaning. This greatly simplifies our understanding of the nature of the exchange / transfer of this quantities an interaction making a strict global conservation possible.

\end{document}